\documentclass[12pt,tightenlines,eqsecnum,floats,shownopacs,nofootinbib,amsmath,amssymb,aps,prd]{revtex4}
\pdfoutput=1 
\usepackage{amssymb}
\usepackage{amsmath,amssymb,amsfonts,amsthm}
\usepackage{graphicx}

\usepackage{setspace}
\usepackage[colorinlistoftodos]{todonotes}
\usepackage{epstopdf}

\def\be{\nopagebreak[3]\begin{equation}}
\def\ee{\end{equation}}
\def\ba{\nopagebreak[3]\begin{eqnarray}}
\def\ea{\end{eqnarray}}

\def\lp{\ell_{\rm Pl}}

\def\f{\frac}

\def\LQC{\rm LQC}

\def\B{{\rm B}}

\def\t{\tilde}
\def\h{\hat}

\def\x{\vec{x}}

\def\R{\mathcal{R}}

\def\U{\mathfrak{A}}

\def\T{\mathcal{T}}

\def\R{\mathcal{R}}
\def\Q{\mathcal{Q}}

\begin{document}

\title{Loop quantum cosmology:\\ From pre-inflationary dynamics to observations}
\author{Abhay Ashtekar}
\email{ashtekar@gravity.psu.edu}
\affiliation{Institute for Gravitation and the Cosmos \& Physics
  Department, Penn State, University Park, PA 16802, U.S.A.}
\author{Aur\'elien Barrau}
\email{Aurelien.Barrau@cern.ch}
\affiliation{Laboratoire de Physique Subatomique et de Cosmologie, Universit\'e Grenoble-Alpes, CNRS/IN2P3\\
53 Avenue des Martyrs, 38026 Grenoble, France}

\begin{abstract}

The Planck collaboration has provided us rich information about the early universe, and a host of new observational missions will soon shed further light on the `anomalies' that appear to exist on the largest angular scales. From a quantum gravity perspective, it is natural to inquire if one can trace back the origin of such puzzling features to Planck scale physics. Loop quantum cosmology provides a promising avenue to explore this issue because of its natural resolution of the big bang singularity. Thanks to advances over the last decade, the theory has matured sufficiently to allow concrete calculations of the phenomenological consequences of its pre-inflationary dynamics. In this article we summarize the current status of the ensuing two-way dialog between quantum gravity and observations.

\end{abstract}


\maketitle

\section{Introduction}
\label{s1}

It is a striking fact that, using known physics, one can account for the observed large scale structure of the universe starting from the tiny  homogeneities in the cosmic microwave background (CMB) as seeds. The presence of these very specific seeds, in turn, can be successfully accounted for in the inflationary scenario, which refers to a much  earlier epoch when matter density and curvature of the universe were over $100$ orders of magnitude higher than they were in the CMB era. This leap is impressive. Furthermore, the underlying paradigm is conceptually attractive because it reduces the issue of genesis to \emph{vacuum fluctuations} at the onset of inflation.
 
Although the curvature scale of inflation is so high compared to that of the CMB epoch, it is still significantly lower than the Planck scale. Therefore, it is safe to assume that space-time geometry is well-described by general relativity (GR) during and after inflation.  This is a happy circumstance, allowing one to ignore what happened before. But from a deeper perspective, this is only a stopgap measure. For, if one goes further back in time, we encounter the Planck regime where quantum field theory (QFT) in curved space-times used in standard inflation is no longer applicable \cite{brandenberger}. We need quantum gravity. Now, there is a long-standing expectation that quantum gravity effects will resolve the big bang singularity of GR. This would require \emph{very large} quantum corrections to the underlying geometry, resulting in a paradigm shift at the Planck scale. One is therefore led to ask: Will inflation arise naturally in this deeper theory? Or, more modestly, can one at least obtain a consistent quantum gravity extension of the inflationary scenario? Can one meaningfully specify initial conditions in the Planck regime? In a viable quantum gravity theory, this should be possible because there would be no singularity and the Planck scale physics would be well-controlled. Would the resulting systematic evolution from the Planck epoch again lead to the correlation functions and the spectral index that are compatible with observations? If not, that quantum gravity approach would be \emph{ruled out} at least in the cosmological sector. If these CMB features are consistent with observations, are there new predictions for future missions which keep memory of the pre-inflationary dynamics? If so, one would be able to directly confront that quantum gravity theory with observations. Thus, attempts to overcome conceptual incompleteness of the inflationary scenario can provide novel ways to test and guide candidate quantum gravity theories.

Thanks to systematic investigations over the past decade, loop quantum cosmology (LQC) is now sufficiently developed to address these issues. As is common in physics, a more fundamental analysis introduces a new scale at which novel phenomena can occur. In LQC, as we discuss in section 2, because of the underlying \emph{quantum} geometry, the big bang singularity is resolved and replaced by a quantum bounce. The curvature at the bounce is universal and introduces a new length scale $\ell_{\LQC}$. The key new phenomenon is the following: Pre-inflationary LQC dynamics modifies the standard inflationary predictions in a universal way for modes whose wavelength at the bounce is larger than $\ell_{\LQC}$. Detailed analysis shows that these correspond either to the longest wavelength modes observable today and/or modes whose wavelength is larger than the radius of the observable universe but which  can couple to the observable modes \cite{schmidt-hui}. Therefore the pre-inflationary dynamics of LQC can have interesting ramifications for the $\sim 3\sigma$ anomalies in the Planck data associated with the largest angular scales.

At first reading, this assertion may seem counter-intuitive on two accounts. First, one generally expects quantum gravity effects to modify only the short-distance behavior. How could they have any implications to predictions for the longest wavelength modes? Second, it is often claimed that while quantum gravity effects may be conceptually interesting, they will not be relevant for cosmological observations because they will all be diluted away during inflation. We will now discuss why these expectations are \emph{not} borne out.

\begin{figure}[htbp]
\begin{center}
\includegraphics[width=16cm]{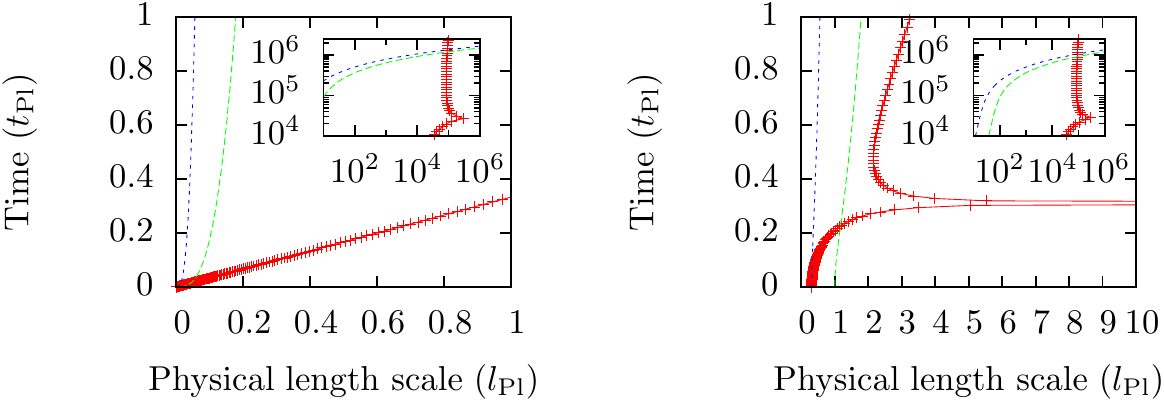}
\caption{\label{fig1} {\footnotesize{Time evolution of the curvature radius shown with (red) solid line with dashes and of two wavelengths of interest to observations, shown with (blue and green) dashed lines from \cite{aan3}.
\emph{Left Panel: General relativity.} The modes of interest have wave lengths less than the curvature radius ($\sqrt{6/R}$, with $R$ is the scalar curvature) all the way from the big bang (${t}=0$) until after the onset of slow roll (${t} \sim 10^{6} t_{\rm Pl}$), shown in the inset.
\emph{Right panel: LQC.} The bounce occurs at time $t=0$ and we have set the scale factor ${a}|_{{t}=0} =1$. The (blue) dotted line on extreme left shows the evolution of the mode whose wavelength $\lambda_{\rm phy}\mid_{{t}=0}$ at the big bounce equals the curvature radius. This mode and modes with smaller wavelengths again remain within the curvature radius until the onset of inflation. On the other hand, modes with physical wavelength \emph{greater} than the curvature radius at
the bounce (for example (green) dashed line) can be excited due to
curvature while their wavelength is greater than the curvature radius and will not be in the BD vacuum at the onset of inflation. (There are two points (at ${t}\approx 0.3\, t_{\rm Pl}$ and ${t}\approx 5\times 10^4\, t_{\rm Pl}$) in the LQC evolution at which the $w=1/3$ in an effective equation of state, whence the scalar curvature vanishes and the radius of curvature goes to infinity.)}}} 
\end{center}
\end{figure}

The belief that the pre-inflationary dynamics does not matter stems from the following argument (see the left panel of FIG. \ref{fig1}). If one evolves the modes that are seen in the CMB \emph{back} in time using GR, their physical wavelengths $\lambda_{\rm phy}$ continue to remain smaller than the curvature radius $R_{\rm curv}$ all the way to the big bang. The equations governing the evolution of these modes then imply that they propagate as though they were in flat space-time and cannot get excited in the pre-inflationary stage. Therefore, the argument goes, they will be in the Bunch-Davies (BD) vacuum at the onset of inflation. 

But in the pre-inflationary calculations, dynamical equations of GR cannot be trusted in the Planck regime; we must use instead a candidate quantum gravity theory. In LQC, if a mode has $\lambda_{\rm phy} > \ell_{\LQC}$ at the bounce, it \emph{does} experience curvature during pre-inflationary dynamics and can get excited (see the right panel of FIG. \ref{fig1}). For suitable choices of initial conditions at the bounce, these modes correspond to the largest angular scales seen in the CMB, roughly to $\ell \lesssim 30$ in the spherical harmonics decomposition of correlation functions.  Thus, the \emph{ultraviolet} modifications of the \emph{background dynamics} that cure the big bang singularity can directly influence the \emph{infrared} behavior of \emph{perturbations.} These longest wavelength modes, then,  will not be in the BD vacuum at the onset of inflation \cite{aan1,aan3}. But why will this fact alter the observable predictions of inflation? Will not these excitations just get washed away during inflation? The answer is in the negative because of the accompanying \emph{stimulated} emission. Agullo, Navarro-Salas and Parker have shown that if one were to start with a candidate non-BD vacuum at the onset of inflation, the stimulated particle creation would result in certain departures from the standard predictions based on the BD vacuum \cite{agulloetal}. The pre-inflationary dynamics of LQC provides specific non-BD initial states at the onset of inflation, thereby streamlining the possibilities and leading to an interplay between the Planck scale physics and observations.

Although loop quantum gravity (LQG) is still far from being complete, given a physical problem, one can carry out a truncation adapted to that problem, and then use the quantum geometry underlying LQG to construct a quantum theory of that sector. This is a strategy commonly used in other areas of physics. For cosmology of the early universe, in the classical theory one uses 
the Friedmann, Lema\^itre, Robertson, Walker (FLRW) geometries, supplemented by first order perturbations on these backgrounds. In LQC one constructs the corresponding quantum theory, using LQG techniques. In section \ref{s2}, we describe the resulting \emph{quantum} FLRW geometries and in sections \ref{s3} and \ref{s4} we discuss cosmological perturbations and the ensuing interplay between theory and observations. Section \ref{s5} contains a summary, a brief description of work in progress, and a few illustrative aspects of the interplay that could not be covered in the main text. Unless otherwise stated, all quantities are in natural Planck units.

The main ideas that refer to developments prior to 2014 are discussed in greater detail in the review articles \cite{asrev,bcgmrev}. 

\section{The quantum FLRW geometry}
\label{s2}

This section is divided into two parts. In the first, we discuss the resolution of the big bang singularity due to quantum geometry effects of LQG. In the second, we summarize the detailed work showing that a phase of slow roll inflation, compatible with observations, arises generically in LQC.

\subsection{Singularity resolution}
\label{s2.1}
Every expanding FLRW solution of GR has a big bang singularity if matter satisfies the standard energy conditions. However, once space-time curvature enters the Planck scale, one expects Einstein's equations to break down. It has been a cherished hope, since 1970s, that quantum effects would step-in and resolve the singularity. Thus, singularities of general relativity --especially the big bang-- are the gates for physics beyond Einstein. Since gravity is encoded in geometry in GR, it is natural to expect that a quantum theory of gravity will require or lead to an appropriate quantum generalization of Riemannian geometry. LQG has taken this premise seriously and systematically constructed a specific theory of quantum  geometry rigorously (for reviews, see, e.g., \cite{alrev,crbook,ttbook}). This theory brings out a fundamental discreteness in the Planck scale geometry. In particular, one can naturally define self-adjoint operators representing geometric observables --such as areas of physical surfaces and volumes of physical regions \cite{al5,al-vol}. It has been shown that geometry is quantized in the direct sense that eigenvalues of geometric operators are discrete \cite{rs,al5,al-vol}. In particular, there is a smallest non-zero eigenvalue of the area operator, which represents the fundamental \emph{area gap}, $\Delta$, of the theory. This \emph{microscopic} parameter sets the scale for new phenomena. 

Specifically, recall that the  Friedmann equation serves as the Hamiltonian constraint that generates the FLRW dynamics on phase space of classical GR. In LQC, the quantum Friedmann equation turns out to be a difference equation, where the step-size is dictated by $\Delta$ \cite{aps}. Recall that in the theory of superconductivity, the energy-gap $\Delta_{\rm E}$ --the energy needed to break a cooper pair-- serves as the microscopic parameter and determines the values of macroscopic parameters such as the critical temperature at which superconductivity sets in: $T_{\rm crit} = ({\rm const})\,\, \Delta_{\rm E}$. The situation is analogous in LQC. Detailed calculations show that the matter density $\rho$ is bounded above, and the value of this macroscopic parameter $\rho_{\rm sup}$ is determined by the microscopic parameter $\Delta$: $\rho_{\rm sup} = ({\rm const})/ \Delta^{3} \approx 0.41 \rho_{\rm Pl}$.%
\footnote{The numerical factor $0.41$ refers to the most commonly used value of the area gap computed using black hole entropy considerations and could be slightly modified by more sophisticated considerations. In comparison with observations, this would alter the value of the LQC parameter $\phi_{B}$ discussed below but not affect any of the main conclusions.} 
 Note that if we let the area-gap $\Delta$ to go to zero  --i.e., ignore the quantum nature of geometry underlying LQG--  $\rho_{\rm sup}$ diverges, quantum effects disappear, and we are led back to the big bang of GR. This is analogous to the fact that, as the the energy gap $\Delta_{\rm E}$ goes to zero, the critical temperature goes to zero and we no longer have the novel phenomenon of superconductivity.
 \begin{figure}
\begin{center}
\includegraphics[width=4.7in,height=2.7in]{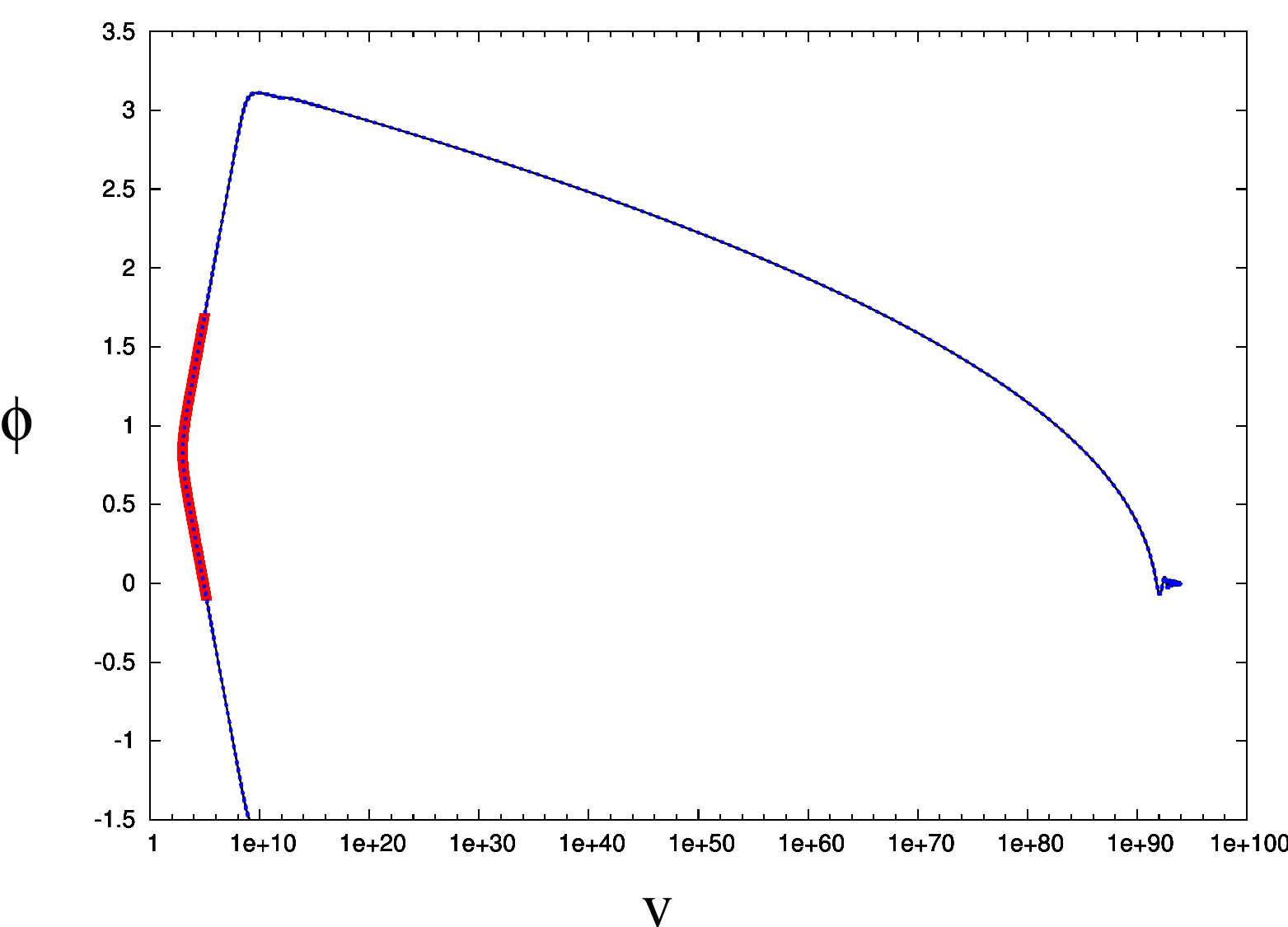}
\caption{\label{fig2} {\footnotesize{An effective LQC trajectory in presence of an inflation with a quadratic potential $(1/2) m^2\phi^2$. Here $V \sim a^3$ is the volume of a fixed fiducial region. The long (blue) sloping line at the top depicts slow roll inflation. As $V$ decreases (from right to left), we go back in time and the inflaton $\phi$ first climbs up the potential, then turns around and starts going descending. In classical general relativity, volume would continue to decrease until it becomes zero, signaling the big bang singularity. In LQC, the trajectory bounces and volume never reaches zero; the entire evolution is non-singular.}}}
\end{center}
\end{figure}
In LQC, then, the big bang singularity is resolved in the following precise sense:  physical observables, such as energy density and curvature which diverge at the big bang in GR, have a \emph{finite} upper bound on the entire Hilbert space of states $\Psi_{o}$ of the FLRW quantum geometry of LQC. (This resolution has also been understood in detail in the `consistent histories' framework \cite{consistent}.) In every physical state, the expectation value of matter density achieves a maximum value $\rho_{\rm max} < \rho_{\rm sup}$, at which the universe bounces, avoiding the formation of a singularity. States $\Psi_{o}$ for which $\rho_{\rm max} \approx \rho_{\rm sup}$ are sharply peaked on trajectories satisfying certain \emph{effective}, quantum corrected equations. In particular, in place of the Friedmann equation $(\dot{a}/a)^{2} = 8\pi G \rho$ of GR, they satisfy \cite{aps}
\be \big({\dot{a}}/{a}\big)^{2}\, =\, 8\pi G \rho\,\, \big(1 -{\rho}/{\rho_{\rm sup}}\big)\, . \ee
At the bounce the right side vanishes and $\dot{a}$ changes sign. These effective trajectories are extremely well-approximated by the classical FLRW solutions for energy densities $\rho \lesssim 10^{-3}\rho_{\sup}$. However, in the Planck regime, there are significant departures. In effect, quantum geometry introduces \emph{a novel repulsive force} which is completely negligible once the curvature is below $\sim 10^{-3}$ in Planck units but which rises very rapidly above this scale and becomes so strong as to overwhelm  the classical attraction and cause the universe to bounce. In the spatially flat, k=0 FLRW models, as one goes back in time from the bounce, the universe expands again and dynamics quickly becomes well approximated by Einstein's equations. FIG. \ref{fig2} shows a bouncing effective trajectory for the the k=0 FLRW universe sourced by an inflaton $\phi$ in a quadratic potential. 

By now, a large number of cosmological models have been studied in detail in LQC, including the closed and open FLRW models, models with a cosmological constant, the Bianchi models and the Gowdy models which incorporate the simplest types of inhomogeneities in full GR. Detailed investigations were carried out using Hamiltonian methods and canonical quantization, complemented by a sum over histories analysis \`a la Feynman for FLRW and Bianchi I models \cite{pathintegral}. In all cases, the singularity is resolved. (This is notable already for Bianchi models where, because the anisotropic shear terms grow as $1/a^{6}$ near the big bang in GR, singularity resolution has been difficult in other approaches.) A general pattern that has emerged can be summarized as follows. GR is an excellent approximation to LQC at low curvature. But when curvature grows to $\sim 10^{-3}$ in Planck units, a novel repulsive force originating in quantum geometry kicks in. It has a `diluting effect' that counteracts the continued growth of curvature that would have occurred in GR. That repulsive forces with origin in quantum mechanics can have macroscopic, even astronomical, implications is already known: for example, it is the Fermi degeneracy pressure that balances the tremendous gravitational pull in neutron stars, enabling them to exist. But in LQC, the origin of the repulsive force lies not in the fermionic nature of matter, but rather in specific properties of the LQG quantum geometry. (For a review of singularity resolution in LQC, see, \cite{asrev}.)

To summarize, while there is no general `singularity resolution theorem' as yet, LQC results (together with the Belinskii, Khalatnikov, Liftschitz conjecture in classical GR) suggest that quantum geometry has an in-built mechanism for resolution of strong curvature, cosmological singularities of GR \cite{ps}.

\subsection{Occurrence of a slow roll phase of inflation}
\label{s2.2}

For concreteness, let us consider a single inflation $\phi$ with $V(\phi) = (1/2) m^{2}\phi^{2}$. In the standard inflationary scenario we are led to set $m= 1.21 \times 10^{-6}$ (in Planck units) by comparing the theoretical power spectrum with observations. This fixes the matter content of the FLRW model. As mentioned in section \ref{s2.1}, there exist wave functions $\Psi_{o}$ satisfying LQC dynamical equations which remain sharply peaked at effective trajectories. In these states the bounce occurs at a density $\rho \approx \rho_{\rm sup}$. While the effective trajectories do not have information about quantum fluctuations in these states $\Psi_{o}$, for many questions it suffices to focus just on the effective equations because the states remain so sharply peaked. This is in particular the case for the question of whether a slow roll phase compatible with inflation occurs `generically' in LQC. 

This question is fraught with ambiguities because the notion of genericity requires a normalized measure on the space of dynamical trajectories and the natural candidate --the Liouville measure-- assigns an infinite volume to this space \cite{ambiguities}. However, physically, this infinity is rather spurious because it is associated with the gauge freedom $a \to \lambda\, a$, with $\lambda \in \mathbb{R}^{+}$, in the definition of the scale factor. It is tempting to factor out the gauge orbits and work on the resulting `reduced', physical space. But the problem is that the Liouville measure is not invariant under this rescaling and therefore does not descend to the orbit space. An alternative is to fix this gauge, effectively declaring that $a(t_{o}) =1$ at some specific time $t_{o}$. Then, it is meaningful to ask the question: What is the fraction of the initial data at $t=t_{o}$ which, when evolved will encounter the desired slow roll some time in the future? However, as one would expect, the answer depends critically on the choice of $t_{o}$. If one chooses it sufficiently early, one finds that the fraction is close to 100\% \cite{klm} while if one chooses it sufficiently late, the fraction is suppressed by a huge factor \cite{gt}. And in GR, there is no preferred choice of $t_{o}$. One might think of choosing the Planck time (as in \cite{klm}) but this is difficult to justify because the Einstein dynamics on which this analysis is based cannot be extrapolated to the Planck regime. (For a detailed discussion of the formulation of the question, the Liouville measure, and the resulting ambiguities in GR, see \cite{aads}. That analysis, and the number quoted in this section, are based on WMAP7 data.) 

In LQC, on the other hand, the situation is conceptually different, first  because now the bounce provides the required preferred instance, and second because the effective equations are valid also in the Planck regime. Investigations have been carried out along two different lines by replacing the FLRW solutions with the solutions to the effective equations of LQC. In the first, \cite{aads,ck}, one uses the bounce time $t_{\rm B}$ for $t_{o}$. Because in LQC the total matter density $\rho$ is bounded above, $\rho \le \rho_{\rm sup}$, and the numerical value of $\rho_{\rm sup}$ is approximately $0.41 \rho_{\rm Pl}$, it follows that the value $\phi_{\B}$ of $\phi$ at the bounce is bounded: $|\phi_{\B}| \le 7.47 \times 10^{5}$. It turns out that if $\phi_{B} > 0.93$, the effective trajectory with that initial data at the bounce point will necessarily enter the slow roll phase \emph{compatible with observations}. In this sense, the fraction of the initial data at $t_{\B}$ that does \emph{not} enter the desired slow roll phase in the future is less than $1.2\times 10^{-6}$. The second analysis re-enforces the genericity of the desired slow roll within LQC by studying the evolution starting before the bounce \cite{bl}. In that case, based on the symmetries of the system, the phase angle of pre-bounce oscillations was taken as a natural parameter for which one can assume a probability distribution. This distribution was demonstrated to be preserved over time. The result is that a sufficiently long inflation is extremely probable. In addition, the number of e-folds of inflation (and, equivalently, the fraction of potential energy at the bounce) can also be accurately predicted and is compatible with data.

\section{Cosmological perturbations: Quantum fields on quantum FLRW geometries}
\label{s3}

This section is divided into three parts. In the first, we explain the main strategy, in the second we discuss the issue of initial conditions and in the third we summarize results on the interplay between theory and observations.

\subsection{The Strategy}
\label{s3.1}

In standard inflation, cosmological perturbations are represented by quantum fields on a FLRW space-time which satisfies Einstein's equations, with the inflaton serving as the matter source. In LQC, we no longer have a fixed FLRW metric in the background, but only a probability amplitude for the occurrence of solutions to the effective equations, encoded in the wave function $\Psi_{o}(a,\phi)$ \cite{aps,pathintegral}. An immediate consequence is that we no longer have a sharply defined proper or conformal time. This apparent obstacle in specifying dynamics can be overcome by using the inflaton $\phi$ as an `internal' time variable with respect to which the scale factor evolves.%
\footnote{Technically, one deparametrizes the Hamiltonian constraint of LQC. It suffices to carry out this deparametrization only `locally' in the regime in which inflaton is single-valued.}
With this choice, the main strategy can be stated as follows: Regard the quantum fields representing cosmological perturbations ---$\hat\Q$, the Mukahnov-Sasaki scalar mode, and $\hat{\T}_{I},\,\, I=1,2$,\, the two tensor modes---  as evolving w.r.t. $\phi$ on the \emph{quantum} FLRW geometry, encoded in $\Psi_{o}(a,\phi)$. Because this background geometry already incorporates quantum gravity effects, this strategy provides a systematic and natural avenue to face the so called `trans-Planckian issues' squarely. 

The obvious question then is whether one can construct the required QFT on  
\emph{quantum} FLRW geometries. Since QFT in curved space-times is already quite intricate, at first this extension appears to be completely  intractable. However, an unforeseen simplification arises \cite{akl}. Let us suppose that the quantum state $\psi(\Q, \T;\, \phi)$ of perturbations is such that the back reaction of the quantum fields $\hat\Q,\, \hat{\T}_{I}$ on the background quantum geometry $\Psi_{o}(a,\phi)$ is negligible, i.e., that $\hat\Q,\, \hat{\T}_{I}$ can be treated as \emph{test fields}.%
\footnote{The test field approximation is also made in standard inflation. There, as well as in LQC, one justifies it by checking that it does hold in the final solution. Note that one uses $\Q$, rather than the curvature perturbation $\R$, because $\R$ is ill-defiend if $\dot\phi$ vanishes and this does occur in the pre-inflationary phase. To obtain the scalar power spectrum, in LQC one calculates $\R$ from the Mukhanov-Sasaki variable $\Q$ at the end of inflation.}
Then, if we start out with an initial state of the form $\Psi_{o}\otimes\psi$ at the bounce, the evolved state continues to have the same tensor-product form, where $\Psi_{o}$ evolves as though there was no perturbation but the evolution of $\psi$ depends on $\Psi_{o}$. Under this assumption, one can recast dynamics of $\psi$ in a way that is \emph{completely tractable} using existing techniques from QFT in curved space-times \cite{akl,aan1,aan3}. 

For a precise statement of this result, let us begin with the tensor perturbations $\T_{I}$. In the classical theory, these fields satisfy the wave equation $\Box \T_{I} =0$ on the background FLRW geometry. In LQC, one can show that the dynamics of the state $\psi(\T_{I};\,\phi)$ on the quantum FLRW geometry $\Psi_{o}$ is \emph{completely} equivalent to that of a state evolving on a certain effective FLRW metric $\tilde{g}_{ab}$ which is `dressed' by quantum corrections. Therefore, although  $\tilde{g}_{ab}$ is smooth, its coefficients depend on $\hbar$ and, more importantly, have to be constructed from the quantum FLRW geometry $\Psi_{o}(a,\phi)$ in a specific fashion. Detailed  calculations \cite{akl} reveal that $\tilde{g}_{ab}$ is given by:
\be \label{qcg} \tilde{g}_{ab} dx^a dx^b \equiv d\tilde{s}^2 =
\tilde{a}^{2} (-d\tilde{\eta}^{2}\, + \,  d{\x}^2 )\, .\ee
where
\be \label{qpara} 
\tilde{a}^4 = \f{\langle \hat{H}_o^{-\f{1}{2}}\,
\hat{a}^4(\phi)\, \hat{H}_o^{-\f{1}{2}}\rangle}{\langle
\hat{H}_o^{-1}\rangle};\quad \quad
d\tilde{\eta} = \langle \h{H}_{o}^{-1/2}\rangle\, (\langle \h{H}_{o}^{-1/2}\, \h{a}^{4}(\phi)\, \h{H}_{o}^{-1/2} \rangle)^{1/2}\,\, d\phi\,\,. \ee
Here, all quantities refer to the Hilbert space of the background FLRW quantum geometry: the expectation values are taken in the state $\Psi_o$,\, $\h{H}_{o}$ is the `free' Hamiltonian in absence of the inflaton potential,\, and $\h{a}(\phi)$ is the (Heisenberg) scale factor operator \cite{akl,aan3}. To summarize, then, if the test field approximation holds, evolution of $\psi(\T_{I};\,\phi)$ on the quantum geometry $\Psi_{o}$ can be described using QFT of $\h{\T}_{I}$ propagating on the quantum corrected FLRW metric $\t{g}_{ab}$. 

Next, let us consider the scalar perturbation $\h\Q$. In QFT on classical FLRW space-times, it satisfies  $(\Box + \U/a^{2}) \h{\Q} =0$  where $\U$ is a potential constructed from the background FLRW solution. In the LQC dynamics, this $\U$ is also dressed, and is replaced by \cite{aan3}
\be \label{qpot} \t{\U}(\phi) = \f{\langle \h{H}_o^{-\f{1}{2}}\,
\h{a}^2(\phi)\, \h{\U}(\phi) \h{a}^2(\phi)\, \h{H}_o^{-\f{1}{2}}
\rangle}{\langle \hat{H}_o^{-\f{1}{2}}\, \hat{a}^4(\phi)\,
\hat{H}_o^{-\f{1}{2}}\rangle}\, . \ee
Thus, the evolution equation of the scalar mode $\h\Q$ is now given by $(\tilde\Box + \tilde\U)\, \h{Q} = 0$. Note that the scalar modes `experience' the same effective metric $\t{g}_{ab}$ as the tensor modes. It is evident from (\ref{qpara}) and (\ref{qpot}) that the expressions of the dressed effective metric and the dressed effective potential could not have been guessed a priori. They resulted from explicit, detailed calculations \cite{akl,aan3}. They `know' not only the effective trajectory on which $\Psi_{o}$ is peaked but also certain fluctuations in $\Psi_{o}$. Finally note that, under the test field approximation, the equivalence is \emph{exact}; it does not involve any additional assumptions, e.g., on the wavelengths of the modes. This is rather striking.\\

\emph{Remark}: The following analogy is helpful in making the main result intuitively plausible. Consider light propagating in a material medium. Photons have a complicated interaction with the molecules of the medium. But so long as they do not significantly affect the material itself ---i.e., so long as the test field approximation holds for photons--- their propagation is well-described by just a few parameters such as the refractive index and birefringence which can be computed from the microscopic structure of the material. In our case, the quantum geometry encoded in $\Psi_{o}$ acts as the medium and quantum perturbations $\h{Q},\h{\T}_{I}$ are the analogs of photon field. So long as the test field approximation holds, the propagation is not sensitive to all the details  of the state $\Psi_{o}$. It can be described using just three quantities, $\t{a},\, \t\eta$ and $\t{\U}$ that are extracted from $\Psi_{o}$ via (\ref{qpara}) and (\ref{qpot}).
\\

This result provides a natural strategy to analyze the pre-inflationary dynamics in LQC by proceeding in the following steps. \emph{(1)} Select a state $\Psi_{o}$ that is sharply peaked on an effective trajectory, which in turn is determined by the value $\phi_{\B}$ of the inflaton at the bounce.  $\phi_{\B}$ turns out to be the \emph{new parameter} that determines the physical wavelengths at the CMB epoch for which LQC effects are important.  \emph{(2)} Calculate the dressed, effective metric $\t{g}_{ab}$ and potential $\t{\U}$ starting from $\Psi_{o}$. \emph{(3)} Select initial conditions for the quantum state $\psi$ of perturbations at the bounce.  \emph{(4)} Evolve this state using QFT on the dressed FLRW metric $\t{g}_{ab}$ and calculate the power spectrum and other correlation functions at the end of inflation. \emph{(5)} Check if the test field approximation holds throughout the `quantum geometry regime'. If not, one has to discard the solution. But if the approximation does hold, then $\Psi_{o}\otimes\psi$ would be a \emph{self-consistent solution}, representing an extension of standard inflation over the 11 orders of change in curvature that separate the onset of inflation from the Planck scale. \emph{(6)} Finally, check the \emph{physical} viability of the solution and compare the predictions of power spectra and correlation functions with observations. The discussion of the next two sub-sections will show that, by now, LQC has matured sufficiently to complete all these steps. 

\subsection{Initial conditions}
\label{s3.2}

Conceptually, the most non-trivial open issue is that of selecting initial conditions for the quantum state $\psi$ of cosmological perturbations. In standard inflation, these are specified at the onset of the slow roll. Now, during the slow roll, the background FLRW geometry can be approximated by the de Sitter metric and there is a unique regular state ---the BD vacuum--- that is invariant under the full de Sitter group. Therefore, the initial state is assumed to be the BD vacuum.%
\footnote{Since the Hubble parameter is not strictly a constant but slowly decreases during inflation, the actual FLRW geometry does not really have de Sitter symmetries. Therefore, conceptually, there is an ambiguity in this choice which however is not important for phenomenology. Similarly, there is an infrared difficulty that can be ignored in practice.}
By contrast, during pre-inflationary dynamics, the Hubble parameter can change rapidly. To illustrate this difference, let us consider a concrete example with $\phi_{\B} = 1.19$ (used in the numerical simulations discussed below). The bounce is followed by a short superinflationary phase during which the Hubble parameter increases from \emph{zero} to its \emph{maximum} in LQC, $H_{\rm sup} = 0.92$, in just 0.18 Planck seconds. It then takes approximately $10^{6}$ Plank seconds till the onset of slow roll inflation at which time the Hubble parameter has decreased to $H_{\rm onset} = 7.8\times 10^{-6}$. Thus, while the duration between the bounce and the onset of inflation is about the same as that of the slow roll itself, the dressed effective metric $\t{g}_{ab}$ does not \emph{at all} resemble the  de Sitter metric in this pre-inflationary phase (for details, see \cite{aads}). Therefore there is \emph{no} reason to assume that the perturbations are in the BD vacuum during this phase. However, compatibility with observations requires that the state resulting from pre-inflationary evolution should be sufficiently close to the BD vacuum at the onset of inflation. A priori this is a heavy burden on pre-inflationary dynamics. But this requirement can be naturally met in LQC.

While the quantum geometry $\Psi_{o}$ does not have full de Sitter symmetry, it is invariant under the six-dimensional Euclidean group $G_{\rm E}$ of space translations and rotations. Therefore, it is natural to require that the state $\psi$ of perturbations is also \emph{invariant under the $G_{\rm E}$ symmetry}. Next, to check that the test field approximation holds, we need to calculate the stress-energy tensor in the state $\psi$. Fortunately, because of the main result described in section \ref{s3.1}, we can lift the machinery of regularization and renormalization of this operator from QFT in curved space-times to QFT on the quantum FLRW geometry encoded in $\Psi_{o}$. The result is that we need to require $\psi$ to be \emph{regular to 4th adiabatic order} w.r.t the dressed, effective metric $\t{g}_{ab}$ (since the stress-energy tensor is an operator of dimension 4). If these two conditions are imposed initially, they are guaranteed to hold throughout the evolution. The last condition on $\psi$ is motivated by the test field approximation. We only assume that it holds \emph{initially}; whether it continues to hold in the subsequent evolution has to be checked explicitly. Thus, the choice of the state 
$\psi(\Q,\T ;\, \phi)$ is subject to three general constraints: i) symmetry; ii) regularity; and iii) validity of the test field approximation at the \emph{initial instant of time}. However, because the $G_{\rm E}$ symmetry is so much weaker than de Sitter, these conditions do not suffice to select a unique state; a large freedom still prevails.  

At what time should the initial state be specified? One possibility is to choose the bounce-time and another is to choose an instant in the distant past in the pre-bounce phase. Both ideas have been pursued. For concreteness, in this section we will summarize the results that arise from the first choice which is motivated by the following considerations. As Roger Penrose, in particular, has emphasized, to account for the observed homogeneity in the CMB, one has to assume an extraordinary degree of homogeneity in the very early universe \cite{rp1}. Note that, by itself, the near exponential expansion through 55-70 e-foldings of inflation does \emph{not} solve this problem: One can still be left with a large inhomogeneity in the CMB, if one started with correspondingly larger inhomogeneities at the onset of inflation. In LQC, the observable universe expands from a ball of radius less than $10\lp$ at the bounce. Therefore, to arrive at the one part in $10^{5}$ homogeneity observed in the CMB, one needs this ball to be extraordinarily  homogeneous. On the other hand, in the pre-bang, collapsing branch, one would expect the geometry to become increasingly wrinkly and complicated. However, as explained in section \ref{s2.1}, in LQG  a repulsive force arises due to quantum geometry which is very effective in diluting such wrinkles in the Planck regime. This `smoothening effect' appears to occur on the $\sim 8 \lp$ scale. Therefore, it can provide a natural mechanism to iron out inhomogeneities, erasing the memory of the pre-bounce phase on balls of radius less than $10\lp$. This scenario provides a natural justification to specify the initial state at the bounce. More precisely, if we restrict ourselves to these balls at the bounce time, \emph{dynamical} considerations from LQG may naturally lead to an initial state of the form $\Psi_{o}\otimes \psi$ where $\Psi_{o}$ represents a quantum FLRW geometry and $\psi$ a small perturbation thereon. In this section, we will discuss investigations based on this choice. However, we should emphasize that so far this reasoning has remained qualitative and serves only as a motivation ---rather than a watertight argument--- to specify the initial state at the bounce. 

To summarize, the strategy is to choose the state $\psi$ of perturbations at the bounce time satisfying the three conditions specified above. So far the focus been on the issue of \emph{existence} of viable initial states rather than uniqueness. That is, the question has been: Can we choose $\psi\mid_{\phi = \phi_{\B}}$ such that: i) the test field approximation holds throughout the period in which quantum geometry effects are important; and, ii) the temperature-temperature correlations $C_{\ell}^{\rm TT}$ agree with observations within error bars? Detailed investigations have shown that such states do exist \cite{aan3,aabg,ia}. The current work focuses on the following issues. Are there interesting observable consequences for future mission? Are there differences from predictions of standard inflation? If so, a possibilities of testing the Planck scale physics of pre-inflationary dynamics would open up. Finally, are there new physical principles that would narrow down the choice of the initial state only to those that are observationally viable? That is, can observations inform quantum gravity?

\subsection{Results}
\label{s3.3}

One avenue to address these questions is to adopt the following viewpoint. The Planck data shows a significant power suppression for $\ell \lesssim 30$ and the origin of this $\sim 3\sigma$ effect is yet to be satisfactorily understood. It may be due to cosmic variance, or a feature of the late time evolution of the universe ---e.g., the integrated Sachs-Wolf effect \cite{isw1,isw2}--- or may have primordial origin. From a quantum gravity perspective, it is natural to assume optimistically that the origin is primordial and explore consequences. In LQC, then, a concrete strategy is to seek initial conditions for $\psi$ at the bounce satisfying the three viability criteria listed in section \ref{s3.2}, \emph{and} for which  there is power suppression at low $\ell$ compatible with the $\sim 3\sigma$ Planck findings. Such initial conditions do exist. 
\begin{figure}[]
  \begin{center}
\includegraphics[width=3.2in,height=2in,angle=0]{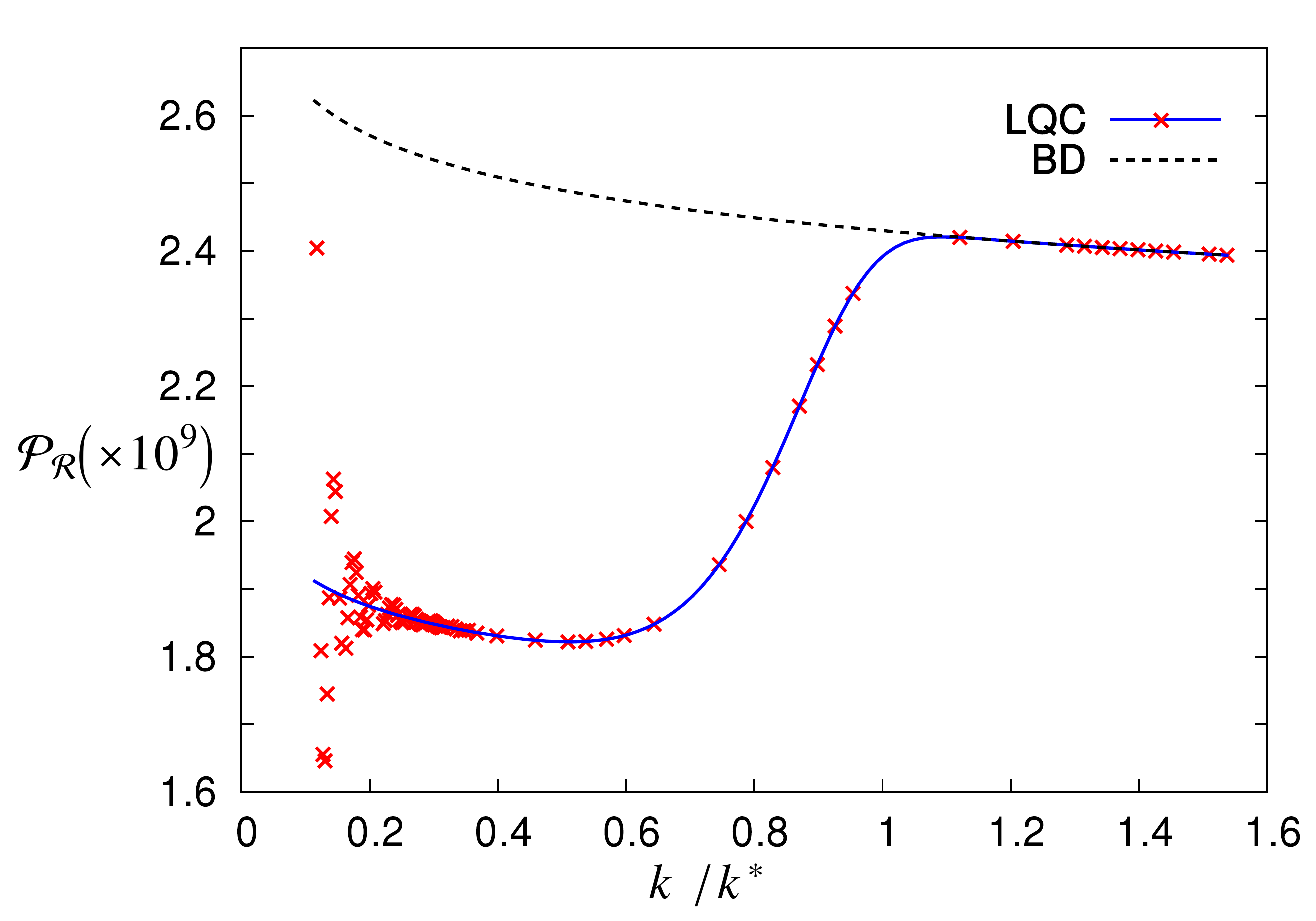}
\includegraphics[width=3.2in,height=2in,angle=0]{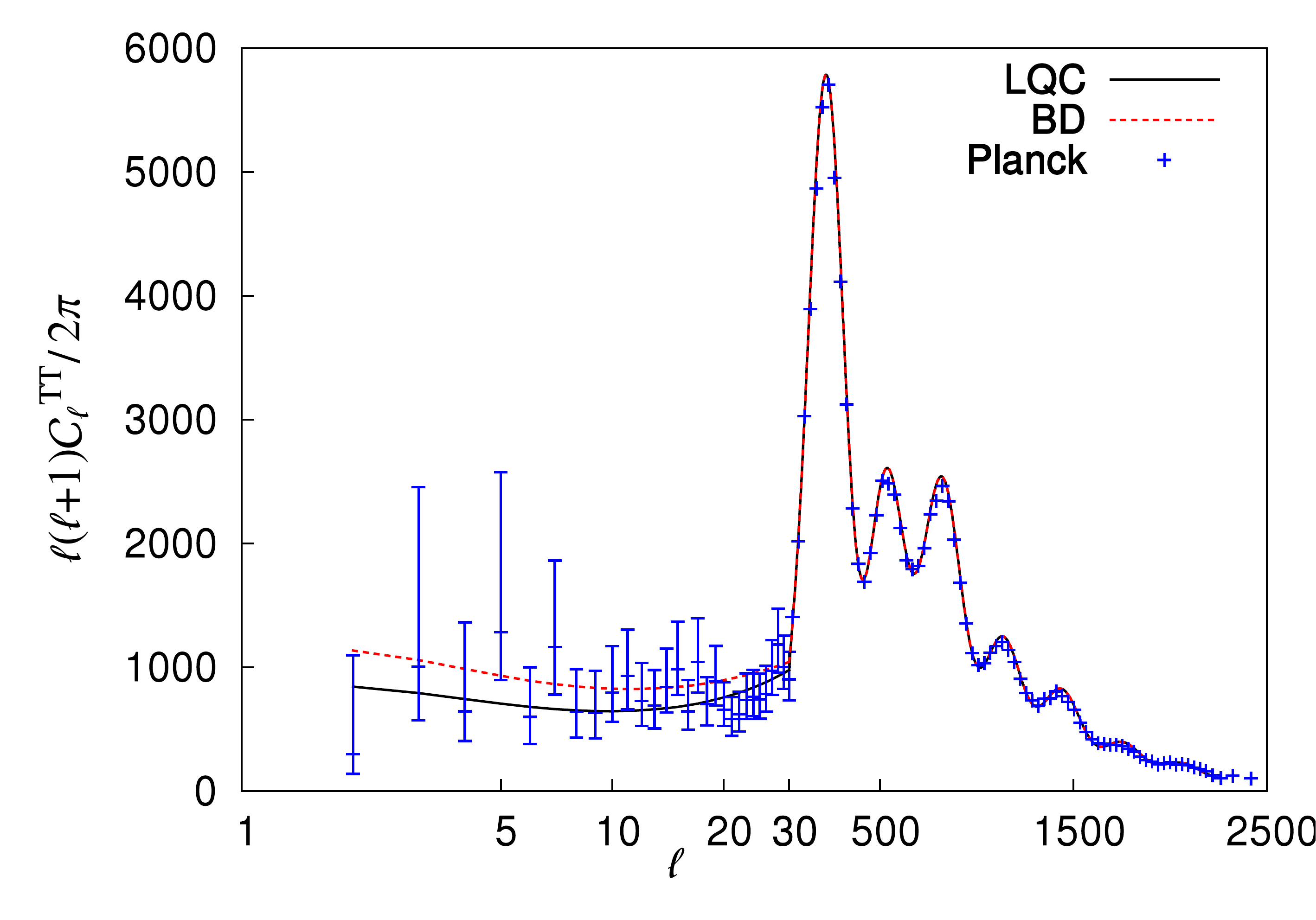}
\caption{\footnotesize{\emph{Left Panel}: Power spectra from LQC and Standard inflation based on the BD vacuum. The (red) crosses are the LQC `data' points, and the solid (blue) line the best fit LQC curve, while the dashed (black) line shows the power spectrum in standard inflation.  $k^{\star}$ is the pivot mode used by WMAP with $k^{\star}/a_{0} = 0.002 Mpc^{-1}$.\,\,\, \emph{Right Panel:} LQC predictions for the T-T correlation function are shown using a solid (blue) line and those of standard inflation, using a (red) dotted line. Both predictions are within the Planck observational errors but the power-suppressed LQC curve provides a better fit to the data. The horizontal axis is enlarged for the $\ell \lesssim 30$ modes because all three curve agree for $\ell >30$. Figures from \cite{aabg}.}}
\label{fig3}
\end{center}
\end{figure}

Let us begin by examining the regime in which deviations from the standard BD predictions can arise. As explained in section \ref{s1}, pre-inflationary dynamics can lead to deviations for the longest wavelength modes since they experience curvature during their pre-inflationary evolution, get excited, and are therefore \emph{not} in the BD state at the onset of inflation. The key question is whether these modes can be seen in the CMB, i.e., if their wavelength is less than the radius of the observable universe. The answer depends on the number $N$ of e-foldings \emph{between the bounce and the onset of inflation.} Detailed analysis shows that, if $N \gtrsim 15$, then even the longest wavelength modes observed today would be inside the curvature radius from the bounce to the onset of inflation. Then (in the simplest scenario) pre-inflationary dynamics would not lead to any observable departures from standard inflation based on the BD vacuum. On the other hand, if the $N \le 15$, some observable modes will cross the curvature radius during the pre-inflationary phase, and not be in the BD vacuum at the onset of inflation. The number $N$ is determined by the pre-inflationary history of the background FLRW geometry, which in turn is governed by the value $\phi_{\B}$ of the inflaton at the bounce. Detailed calculations show that $N <15$ if $\phi_{\B} < 1.2$ \cite{aan3}. On the other hand, as we saw in section \ref{s2.2}, to obtain a sufficiently long slow roll, we need $\phi_{\B} > 0.93$. Consequently, while we can choose initial conditions for $\psi$ so that there is agreement with the Planck data within error bars, for all $\phi_{B} >0.93$, in the simplest LQC scenario there is only a small window, $0.93 \le \phi_{B} \le  1.2$, in which predictions can differ from standard inflation, based on the BD vacuum. Furthermore, these deviations occur only for the longest wavelength modes with $\ell \lesssim 30$.  

The question then is whether one can arrive at the `anomalies' associated with the largest angular scales by restricting ourselves to this window. This is a primary focus of the current investigations in LQC. Because space is limited, we cannot describe all the ideas that are being pursued. Here we will discuss just one choice of the initial conditions $\psi\mid_{\phi=\phi_{\B}}$ to illustrate how the interplay between LQC and observations is shaping up, and summarize another idea in section \ref{s5}. 

\begin{figure}[]
  \begin{center}
\includegraphics[width=3.2in,height=2in,angle=0]{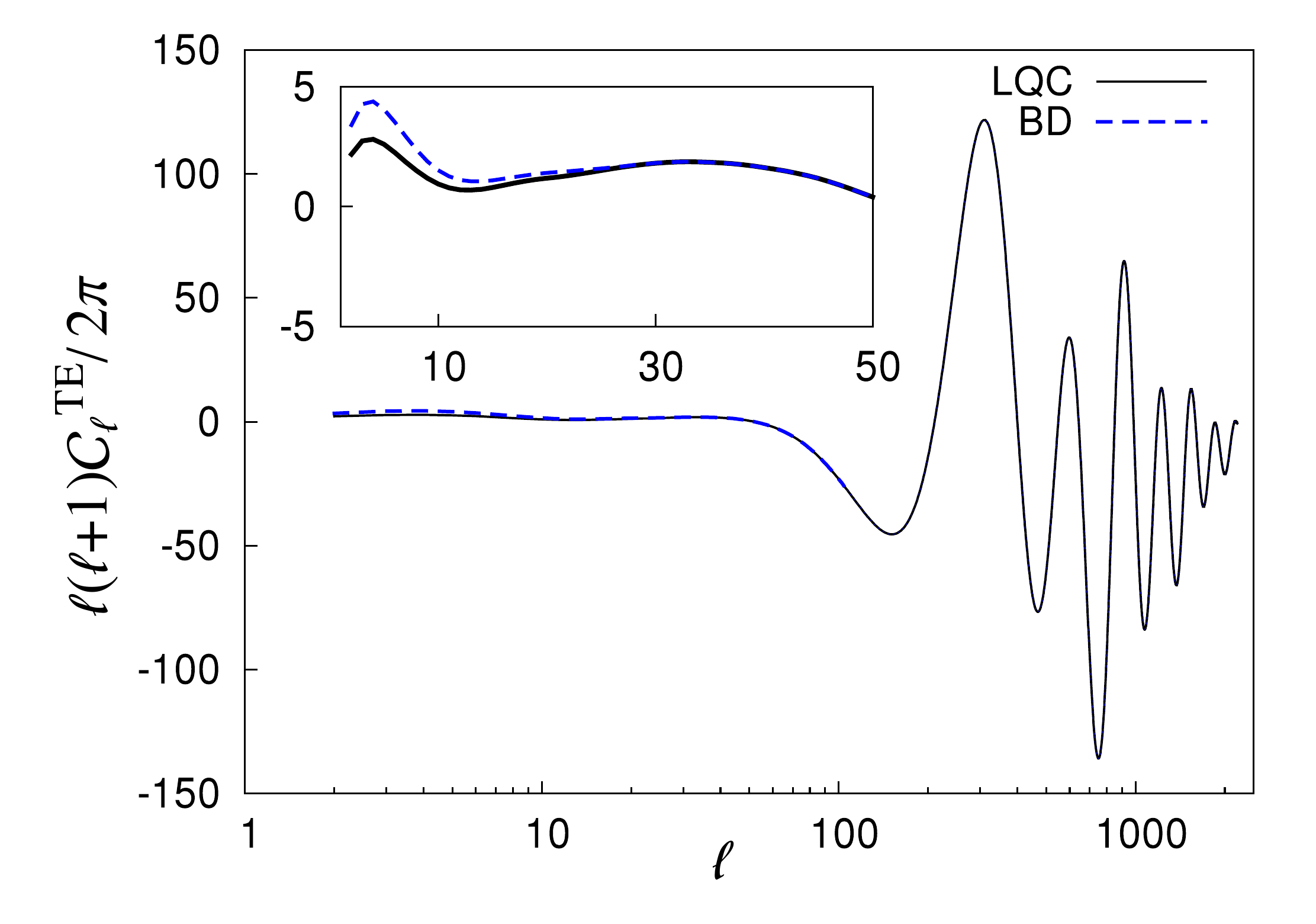}
\includegraphics[width=3.2in,height=2in,angle=0]{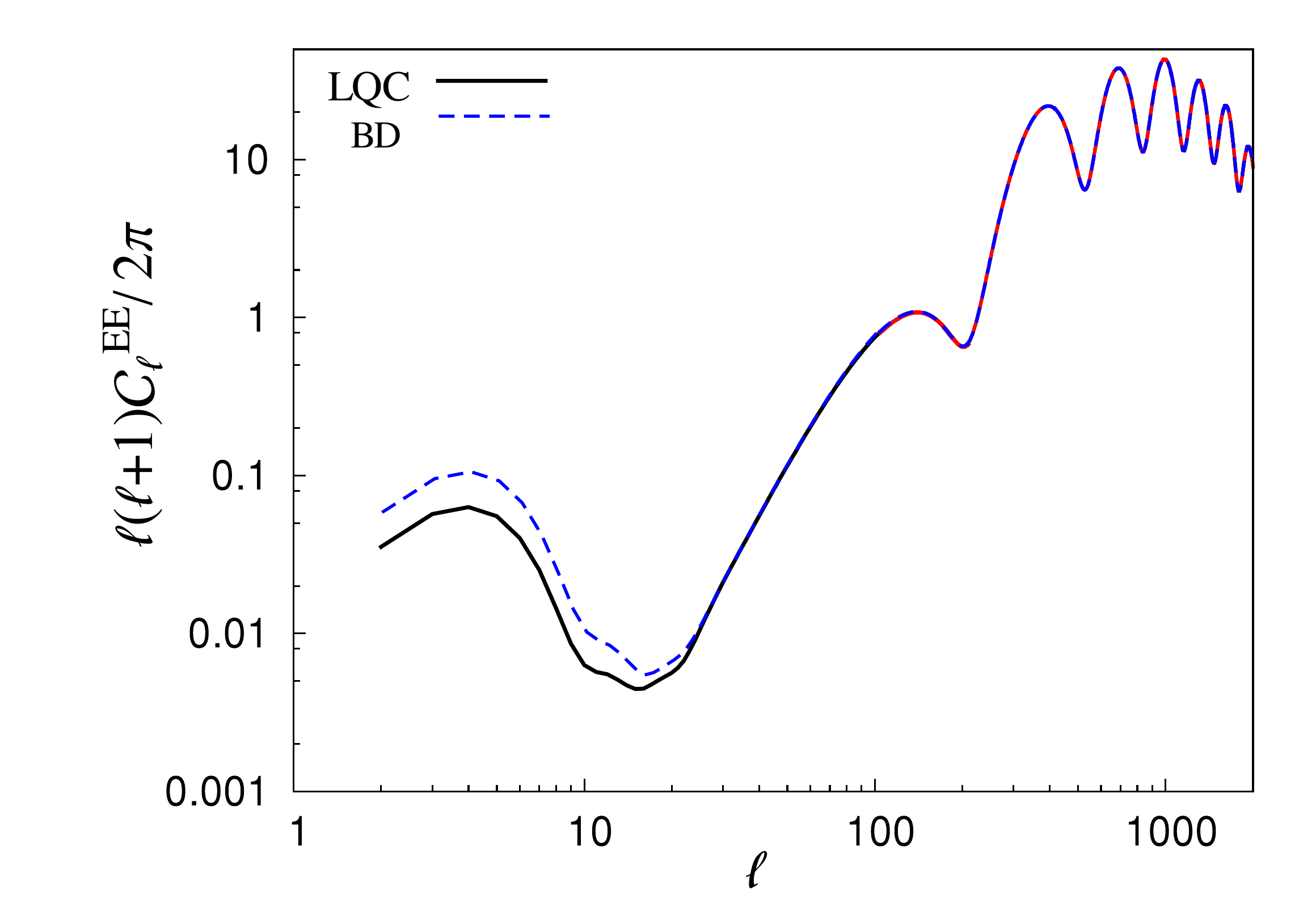}
\caption{\footnotesize{Predictions of LQC are depicted by solid (black) lines and those of standard inflation based on the BD vacuum by dotted (blue) lines for the T-T and the E-E correlations. In \emph{both} cases there is power suppression for $\ell \lesssim 30$ in LQC. Figures from \cite{aabg}.}}
\label{fig4}
\end{center}
\end{figure}

The results \cite{aabg} are shown in FIGs. \ref{fig3} and \ref{fig4}. In this solution for $\Psi_{o}\otimes \psi$, the background quantum geometry state $\Psi_{o}$ is sharply peaked around the effective trajectory determined by $\phi_{\B} = 1.19$, and $\psi\mid_{\phi=\phi_{\B}}$ is such that there is power suppression for the longest wavelengths (i.e., for the lowest co-moving $k$'s). The left panel of FIG. 3 compares the theoretical power spectra for the curvature perturbation $\R$ in LQC and in standard inflation. Here $k^{\star}$ is the WMAP pivot mode,\, $k^{\star}/ a_{o} = 2\times 10^{-3} {\rm Mpc}^{-1}$. The longest wavelength mode observable today corresponds to $k = k_{o}$ with $k_{o} \approx 0.12\, k^{\star}$. The solid (blue) curve is the best fit through the LQC `data points' shown with (red) crosses while the dotted (black) curve shows the prediction from standard inflation. It is clear that the two predictions agree for $k \gtrsim k^{\star}$ but in LQC there is power suppression for observable modes $k$ in the range $k_{o} \lesssim  k \lesssim k^{\star}$. The right panel translates this result to the temperature-temperature correlations $C_{\ell}^{\rm TT}$ for direct comparison with observations. It shows that in LQC there is power suppression relative to standard inflationary prediction for $\ell \lesssim 30$, although both sets of predictions are within the Planck error bars. 

Thus, there are choices of initial conditions for $\psi$ satisfying the three constraints specified in \ref{s3.2} and, in addition, are such that: i) for $\ell >30$, (where the observational error bars are tiny) the resulting power spectrum agrees with that predicted by standard inflation and also with observations; and,  ii) there is desired power suppression for $\ell <30$. The fact that all these conditions are met simultaneously by the LQC pre-inflationary dynamics is non-trivial. But can one do even better? Can this choice of initial conditions be tested through future observations? \emph{The answer is in the affirmative.} One finds that there is also suppression at low $\ell$ for the T-E (temperature-electric polarization) and E-E correlations \cite{aabg}. These are shown in FIG. \ref{fig4}. An interesting aspect of this result is that it serves to distinguish the LQC primordial mechanism from the one tied to the late-time evolution of the universe through the integrated Sachs-Wolf effect \cite{isw2} because, in that case, there would not be power suppression in the E-E correlation functions. Thus, if future observations show that the (T-E and) E-E correlation functions are suppressed relative to the standard inflationary prediction for $\ell <30$, one would have a clear signal in favor of the primordial origin of these `anomalies'. From LQC standpoint, this would be the observable imprint of the Planck scale, pre-inflationary dynamics. 

To summarize, if the LQC parameter $\phi_{\B}$ is in a narrow window ($0.93 \le \phi_{B} \le  1.2$), there exist admissible initial conditions for the state $\psi$ of perturbations such that one can account for the Planck data, including the power suppression at $\ell \lesssim 30$. Furthermore, there are predictions for future missions that distinguish this quantum-gravity based mechanism from non-primordial mechanisms for power suppression. This example illustrates the extent to which LQC has matured in recent years.

\section{Cosmological perturbations: Anomaly-free Hamiltonian dynamics}
\label{s4}

We will now discuss the ``deformed algebra'' approach that focuses on consistency of the effective theory rather than the quantum geometry that underlies LQG.\\

Recall that the constraints of general relativity form a first class system \`a la Dirac and this property is key to the consistency of the classical dynamics. It is not a priori clear whether this delicate consistency can persist in the effective theory that incorporates leading order quantum corrections, in particular in cosmological perturbation equations. Furthermore, in background-independent frameworks, it is not clear whether one can rely on the standard covariance arguments \cite{hkt}, because the very notion of space-time is supposed to emerge in some way from solutions to the fundamental equations. The consistency of the quantum corrected equations must be ensured before they can be solved. The ``deformed algebra" approach puts a specific emphasis on issues of gauge. In other areas of physics, gauge fixing before quantization was shown to be harmless but the case of gravity is much more subtle than, say, Yang-Mills theories because dynamics is part of the gauge. This issue is here taken seriously by trying to build the algebra so that the constraints can be quantized without classical specifications of gauge or observables.\\

In the ``deformed algebra" approach to LQC, one considers two kind of LQG-inspired corrections to the constraints. One is the so-called holonomy correction (coming from the fact that loop quantization is based on holonomies, {\it i.e.} exponentials of the connection, rather than connections themselves), and the other is the inverse-triad correction (coming from terms in the Hamiltonian constraint which cannot be quantized directly but only after being re-expressed as a Poisson bracket). In the following, only holonomy corrections will be considered as their interpretation is easier. The net effect of these corrections is now encoded in the replacement
\begin{equation} \label{replacement}
\bar{k} \rightarrow \mathbb{K}[n] := \frac{\sin(n\bar{\mu} \gamma \bar{k})}{n\bar{\mu}\gamma},
\end{equation}
where $n$ is some unknown integer, $\bar{k}$ is the mean 
gravitational connection, and $\bar{\mu}$ is the coordinate size of a loop. The quantum-corrected constraints resulting from this substitution are denoted by $\mathcal{C}^Q_I$. If the replacement (\ref{replacement}) is performed naively, the algebra reads
\begin{equation}
\{ \mathcal{C}^Q_I, \mathcal{C}^Q_J \} = {f^K}_{IJ}(A^j_b,E^a_i) \mathcal{C}^Q_K+
\mathcal{A}_{IJ},
\end{equation}
where $\mathcal{A}_{IJ}$ stand for anomaly terms and $f^K_{IJ}$ are structure functions. The consistency condition (i.e., closure of the algebra) requires $\mathcal{A}_{IJ}=0$. In turn, very nicely, this condition imposes restrictions on the  form of the quantum corrections. Since the result is the modification of the constraint algebra of general relativity, as one would expect from the Hojman-Kuchar-Teitelboim theorem \cite{hkt}, the quantum space-time structure  can no longer be described in terms of a pseudo-Riemaniann space-time metric \cite{aan2}. But it does have a well-defined \emph{canonical} formulation using hypersurface deformations.\\

The procedure can be described as follows. The quantum corrected constraints are explicitly written for the perturbations up to the desired order. Then the Poisson brackets are calculated. Anomalies are evaluated. Suitable counter-terms, which are required to vanish at the classical limit, are added to the expressions of constraints to ensure the anomaly freedom. The resulting theory is not only consistent but is also -- somehow surprisingly -- quite uniquely defined when matter is included. In addition, although the calculations are quite long and involve intricate expressions, the resulting final algebra is very simple and depends on a unique and elegant structure function which encodes all the modifications :

\begin{equation}
\Omega=1-2\rho/\rho_c. 
\end{equation}

Importantly, the unknown integers $n_i$ (that can, in principle, be different from unity as long as one considers other terms than the simple $\bar{k}^2$ term arising from the curvature of the connection) entering the correction functions  $\mathbb{K}[n_i]$ can {\it all} be determined unambiguously. The solution closes the algebra non-pertubatively. This strategy has been successfully followed for vector \cite{mcbg} and scalar \cite{cmbg} perturbations. It has been shown that a single algebra structure can be consistently written for {\it all} perturbations \cite{cbgv}. It basically reads as:
\begin{eqnarray}
\left\{D[M^a],D [N^a]\right\} &=& D[M^b\partial_b N^a-N^b\partial_b M^a], \\
\left\{D[M^a],S^Q[N]\right\} &=& S^Q[M^a\partial_b N-N\partial_a M^a],  \\
\left\{S^Q[M],S^Q[N]\right\} &=& \Omega 
D\left[q^{ab}(M\partial_bN-N\partial_bM)\right],
\end{eqnarray}
where $D$ and $S$ are the diffeomorphism and hamiltonian constraints, $N$ and $M$ are lapse functions, $N^a$ and $M^a$ are shift functions, $\Omega$ is the ``deformation factor" (equal to one in the classical theory with Lorentzian signature), and the superscript $Q$ indicates that the constraint is quantum corrected at the effective level. 

This algebra is elegant, simple, and presents some interesting features: in particular it leads to an effective signature change, somehow reminiscent of the Hartle-Hawking proposal, close to the bounce. When $\rho<\rho_c/2$ the spacetime structure is Lorentzian but when $\rho>\rho_c/2$, that is in the vicinity of the bounce, $\Omega$ becomes negative and the spacetime structure becomes Euclidean. Strikingly, this effect has also been found independently following different paths in LQC \cite{ed,bp}. In particular, the first of these references relies on very different hypotheses (using a model of patches of universe evolving independently in the longitudinal gauge). The fact that it leads to the same result strongly reinforces the conclusion. At this stage, perturbations in this framework have been analyzed according to two different philosophical viewpoints that we shall now briefly review. \\

It is important to notice that, in any case, the equation of motion is now more complicated than in the ``dressed metric" approach described in section \ref{s3}. For the simplest example of tensor modes, it reads
%
\begin{equation}
		v''_k(\eta)+\left(\Omega k^2-\frac{z_T''}{z_T}\right)v_k(\eta)=0 \label{eom2}
\end{equation}
in conformal time, where  the mode functions $z_T\equiv ({a}/{\sqrt{\Omega}})$  are  related to the amplitude of the tensor modes of the metric perturbation $h_k$, via $v_k=z_T h_k/\sqrt{32\pi G}$. The key point to notice here is that the dynamics of the modes is not anymore driven only by the hierarchy between $k^2$ and $a''/a$ ({\it i.e.}, by the ratio between the length scale associated with the mode and the curvature radius): it is more complicated and quite a lot of new phenomena can appear, opening a wide phenomenology.\\
\begin{figure}[]
  \begin{center}
\includegraphics[width=4in,height=3in,angle=0]{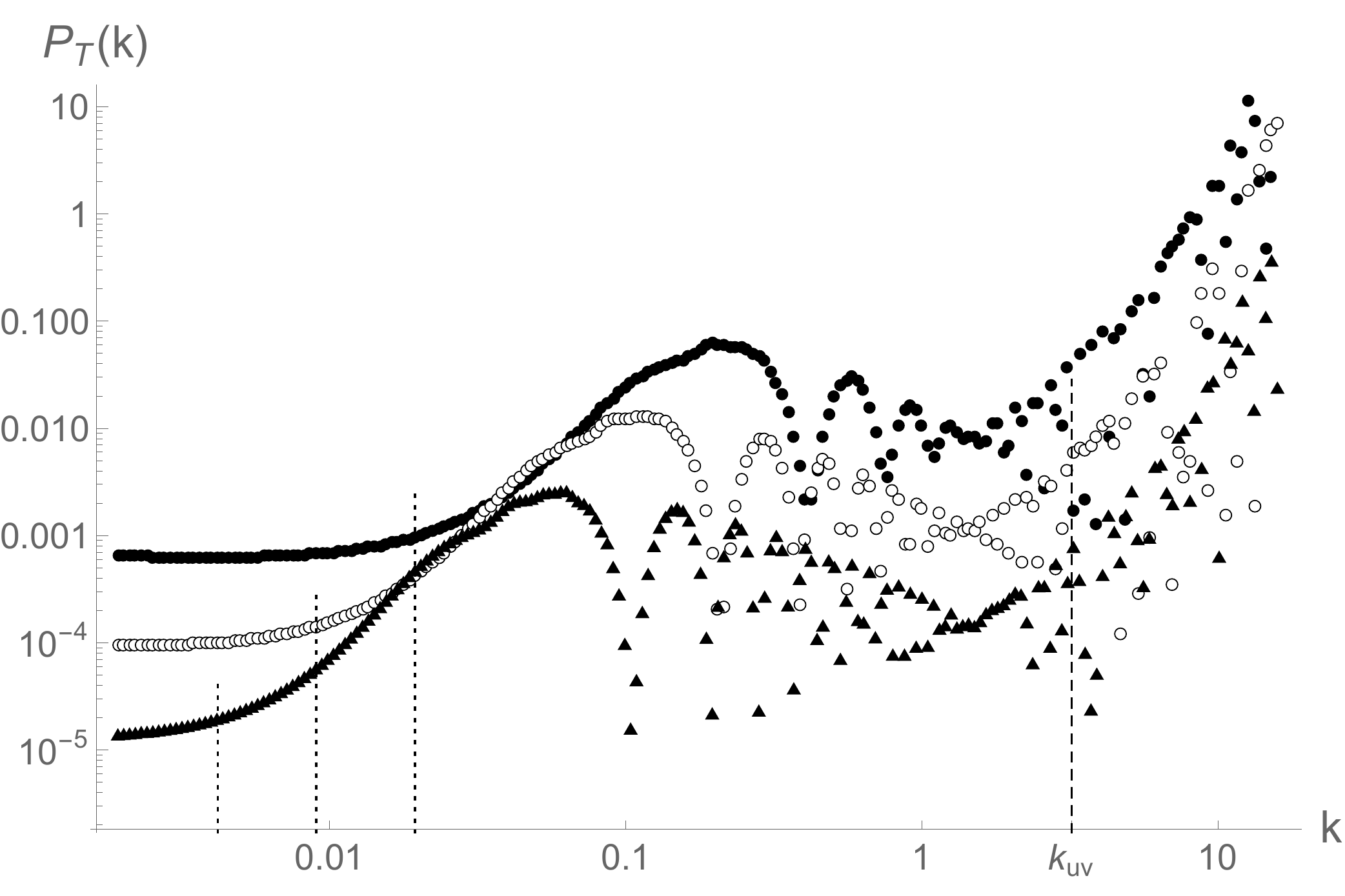}
\caption{\footnotesize{Primordial power spectra for tensor modes in the deformed algebra approach for different values of the mass
of the scalar field: $m = 10^{-3}M_{Pl}$ (triangles), $m = 10^{-2.5}M_{Pl}$ (open disks), and $m = 10^{-2}M_{Pl}$ (black disks). The behavior remains qualitatively similar for smaller (favored) values of the mass (from \cite{bgslb}).}}
\label{fig5}
\end{center}
\end{figure}
The first viewpoint is to use the ``silent surface"corresponding to $\Omega=0$  (or,  $\rho=\rho_c/2$) as the natural place for imposing initial conditions. This is a tantalizing hypothesis as space points are here decoupled (as anticipated by the BKL conjecture which is here nicely demonstrated in this specific quantum gravity setting) and fluctuations are naturally described by a white noise spectrum. The resulting cosmological power spectrum was derived in \cite{mlb} and specific issues were studied in \cite{bm}.

The second viewpoint is to propagate the perturbations through the phase $\Omega<0$ (when the hyperbolic equation becomes elliptic), setting initial conditions in the contracting branch of the universe. This is technically possible. The vacuum is well defined: far enough in the past, all modes are within the horizon and it makes sense to set initial conditions ``earlier" in time and as far as possible from the highly quantum region. The resulting power spectrum was derived in \cite{lcbg,bgslb}. It exhibits an infrared behavior in agreement with GR, an exponentially diverging UV limit%
\footnote{This in itself need not be a problem. First, because the primordial power spectrum should not be trusted up to arbitrary small scales: non-linear physics inevitably enters the game for very high values of $k$. Second, because back-reaction should then be taken into account. This just gives the general trend of the behavior.}  
and oscillations in-between. Those oscillations can easily be interpreted as coming from a potential well in the Shr\"odinger-like equation describing the evolution of perturbations. Figure \ref{fig5} shows the kind of primordial spectra expected in this framework.\\

The important point for detection is the number of e-folds that take place between the bounce and the contemporary universe. For most solutions, the number is very high and the relevant modes are trans-planckian at the bounce (this is true both for the ``dressed metric" and for the ``deformed algebra" approaches). However, for a restricted set of initial conditions, the observational window does fall in the non-trivial part of the spectrum and has interesting features that might be discovered by future experiments devoted to the polarized CMB.

\section{Discussion}
\label{s5} 

That quantum geometry effects of LQC can naturally lead to a resolution of the  big-bang singularity was first realized over a decade ago \cite{mb1}. A large body of subsequent work, summarized in \cite{asrev}, has significantly sharpened the key ideas and made them robust. The community also has detailed results on the generic emergence of a slow roll inflationary phase for the $(1/2)m^{2}\phi^{2}$ and the Starobinsky potentials. Other potentials and multi-inflaton models have also been considered and the qualitative results are the same. Thus, on the whole, there is broad consensus in the LQC community on the results reported in section \ref{s2}. 

Investigations of cosmological perturbations within LQC are more recent.  In all approaches, one focuses on the \emph{`cosmological sector'} of classical general relativity, consisting of the FLRW solutions with inflation as matter source, together with the first order perturbations. This subject has been pursued along related but technically different directions (see, e.g., \cite{aan1,aan3,cbgv,bgslb,madrid}). These investigations provide a healthy diversity. In this brief overview, we could only include illustrative examples from this growing body of results. In section \ref{s3} we discussed the most developed of these approaches. Here, perturbations are treated as quantum fields propagating on background, FLRW \emph{quantum} geometries discussed in section \ref{s2}. Self-consistency of solutions is checked through detailed calculations of the back reaction, and a large number of observational consequences have been systematically worked out. In the second approach, summarized in section \ref{s4}, the viewpoint is rather different: The goal is to capture the leading order quantum corrections that would arise in \emph{any} consistent Hamiltonian quantization of the cosmological sector that is based on connection variables. The idea is to encode these corrections in modifications of the GR constraint equations that are subject to the condition that the Poisson algebra of the modified constraints also closes, i.e., is anomaly-free. Recently, phenomenological consequences of the resulting modified equations of motion have been worked out. Interestingly, although the two approaches are conceptually quite different, their predictions agree for a range of intermediate wavelengths, but differ from standard inflation \cite{bgslb}.

At present both approaches have certain limitations. In the first approach,  the true degrees of freedom for the first order perturbations are extracted at the classical level, prior to quantization. While this is an internally consistent procedure, as a check of robustness, it is of interest to first pass to the quantum theory and \emph{then} extract the true degrees of freedom by imposing the full set of quantum constraints for the background as well as perturbations. This check has not been carried out. A second limitation, shared by all LQC approaches to date, is that quantization is carried out through a scheme that is `hybrid' in the following sense. The quantum theory of the background FLRW sector is based on quantum geometry techniques from LQG and they lie at the heart of  singularity resolution. But then, in the resulting quantum geometry, perturbations are quantized \`a la Fock, rather than by using LQG techniques. This strategy is motivated by the expectation that the Fock states of perturbations should provide an excellent approximation of the loop quantized states because the energy density in perturbations is small. But as of now there is no detailed analysis justifying this expectation. For the second approach, one limitation is that the modified GR constraint equations used at the start of the analysis are not systematically derived. Rather, an ansatz is made at the start of the calculation so that the required calculations are tractable. Also, in presence of these modifications, the notion of space-time covariance has to be
revisited and the hope that there may be an underlying non-commutative geometry is yet to be realized. Finally, recent calculations \cite{bgslb} have shown that this `anomaly-free' method leads to an exponentially growing blue tilt at small angular scales (in contrast to standard inflation, and the approach described in section \ref{s3}, both of which have a small red shift). This blue tilt would lead to a large back reaction and the issue of finding a self-consistent solution remains open in this approach. 

We will conclude with a few examples of work in progress. So far, most of the detailed investigations have been carried out with the  quadratic potential for the inflaton. Other potentials, particularly the one that features in the Starobinsky inflation, are now being analyzed in detail \cite{bbbg}. Results to date already shows that the features reported in section \ref{s3} are robust. Second, so far, the detailed analysis has used states $\Psi_{o}$ for the background geometry that are sharply peaked. Currently more general states in which the uncertainty in the scale factor even higher than $100$\% are being used. Interestingly, more general states do not introduce new parameters vis a vis observations: the power spectrum that results from the widely spread background states is the same as the power spectrum from sharply peaked states, just with a shift in value of the LQC phenomenological parameter $\phi_{\B}$. The origin of this rather surprising effect is now understood in terms of the role played by the microscopic parameter, the area gap $\Delta$, in the theory. More vigorous effort is being made to probe and control the freedom in the choice initial conditions for the quantum state $\psi$ of perturbations. In section \ref{s3} we considered one choice that directly leads to power suppression. But there are other choices \cite{aan3,ana} (also satisfying the three viability criteria of section \ref{s3.2}) for which the power spectrum is enhanced for wavelengths that are larger than the radius of the observable universe. In presence of interactions, correlations between observable modes and these unobservable super-horizon modes produces a modulation in the power spectrum of observable modes. Furthermore, unlike in mechanisms proposed in \cite{schmidt-hui}, the LQC amplitude modulation \emph{decays} at smaller angular scales and is thus compatible with observations. This opens an interesting possibility of obtaining significant non-Gaussian correlations only between the longest wavelength modes we can observe and even longer wavelength modes that we cannot. This mechanism could then be used to account for the observed $\sim 3\sigma$ anomalies at large angular scale \cite{ia}. Furthermore these considerations appear to \emph{reduce} the tensor to scalar ratio $r$, thereby pushing the $(1/2)m^{2}\phi^{2}$ potential to the `allowed region' of the Planck data. Thus, the issue of whether these anomalies can be accounted for using Planck scale physics is being approached from several different directions within LQC. Because the final observational predictions are tied with the choice of initial conditions, these diverse investigations open up the interesting possibility of constraining the initial conditions at the bounce through observations. More precisely, using guidance from observations it may now be feasible to uncover a fundamental principle dictating initial conditions that has eluded us so far. An attractive possibility in this direction is to seek an appropriate quantum generalization of Penrose's Weyl curvature hypothesis \cite{aan3,rp1}. Another possibility is suggested by the ``deformed algebra'' approach along the lines of \cite{mlb}. Such rather diverse investigations are now actively in progress.

In the cosmology literature, it is often implicitly assumed that Planck scale physics would inform the discussion of the early universe mainly by explaining the origin of the inflaton(s) and providing us the potential(s) that dictate their dynamics. Investigations in LQC are silent on these issues because these issues have their origin in particle physics. But LQC has opened another, complementary, window. Quantum geometry underlying LQG introduces specific changes in the dynamics of GR in the ultraviolet that lead to the resolution of the big bang singularity. Interestingly, the new quantum FLRW geometry modifies the dynamics of the longest wavelength modes of perturbations, thereby providing avenues to account for the $\sim 3\sigma$ anomalies seen at the largest angular scale. The resulting two-way interplay between theory and observations has the potential of enriching both.

\section*{Acknowledgments}
We would like to thank Ivan Agullo, Boris Bolliet, Brajesh Gupt and Parampreet Singh for ongoing discussions. AA also benefited from the questions and discussions at Paris conference \emph{Primordial universe after Planck}. This work was supported by the NSF grants PHY-1505411 
and the Eberly research funds of Penn state.


\begin{thebibliography}{99}

\bibitem{brandenberger} R.~H.~Brandenberger, Introduction to early
    universe cosmology, PoS(ICFI 2010)001,
    \texttt{arXiv:1103:2271};\\
    R.~H.~Brandenberger and J.~Martin, Trans-Planckian issues for inflationary cosmology, \texttt{arXiv:1211.6753}

\bibitem{schmidt-hui} F.~Schmidt, L.~Hui, CMB Power Asymmetry from Non-Gaussian Modulation, Phys. Rev. Lett. \textbf{110} 011301 (2013). 

\bibitem{aan1} I. Agullo, A. Ashtekar and W. Nelson, A Quantum Gravity Extension of the Inflationary Scenario,  Phys. Rev. Lett. \textbf{109}  2513 (2012). 

\bibitem{aan3} I. Agullo, A. Ashtekar and W. Nelson, The pre-inflationary dynamics of loop quantum cosmology: Confronting quantum gravity with observations,  Class. Quant. Grav. \textbf{30}  085104 (2013).

\bibitem{agulloetal} I.~Agullo and L.~Parker,  Non-gaussianities
    and the Stimulated creation of quanta in the inflationary
    universe, Phys.~Rev.~D~\textbf{83} 063526 (2011);\\
    Stimulated creation of quanta during inflation and the observable universe Gen.~Relativ.~Gravit.~\textbf{43} 2541-2545 (2011);\\ 
    I.~Agullo, J.~Navarro-Salas and L.~Parker,  Enhanced local-type
    inflationary trispectrum from a non-vacuum initial state, JCAP~ 
    \textbf{1205} 019 (2012).
    
\bibitem{asrev} A.~Ashtekar and P.~Singh, Loop quantum cosmology: A
    status report, Class. Quant. Grav. \textbf{28}, 213001 (2011).

\bibitem{bcgmrev} A.~Barrau, T.~Cailleteau, J.~Grain and J.~Mielczarek, 
Observational issues in loop quantum cosmology, Class. Quant. Grav. \textbf{31} 053001 (2014). 
 
\bibitem{alrev} A.~Ashtekar and J.~Lewandowski, {Background
    independent quantum gravity: A status report}, Class. Quant.
    Grav. {\bf 21} R53-R152 (2004).

\bibitem{crbook} C.~Rovelli,{\em Quantum Gravity}. (Cambridge
    University Press, Cambridge (2004)).

\bibitem{ttbook} T.~Thiemann, {\em Introduction to Modern
    Canonical Quantum General Relativity.} (Cambridge University Press,
    Cambridge, (2007)).

\bibitem{al5}A.~Ashtekar and J.~Lewandowski, Differential geometry
    on the space of connections via graphs and projective limits, J. Geom.
    Phys. \textbf{17} 191-230 (1995).
    
\bibitem{al-vol}A.~Ashtekar and J.~Lewandowski, Quantum theory of
    geometry II: Volume Operators, Adv.~Theo.~Math.~Phys.~\textbf{1} 388-429 (1997). 

\bibitem{rs} C.~Rovelli and L.~Smolin, Discreteness of area
    and volume in quantum gravity, Nucl.~Phys.~B~\textbf{442}
    593--622 (1995); Erratum: Nucl.~Phys.~B~\textbf{456} 753
    (1995).   

\bibitem{aps} A.~Ashtekar, T.~Pawlowski and P.~Singh, {Quantum
    nature of the big bang}, Phys. Rev. Lett. \textbf{96} 141301
    (2006);\\
    {Quantum nature of the big bang: Improved dynamics}, Phys. Rev. D{\bf
    74} 084003 (2006).

\bibitem{consistent} D.~Craig and P.~Singh, Consistent probabilities in loop quantum cosmology, Class. Quantum Grav.\textbf{30} 205008 (2013). 

\bibitem{pathintegral} A.~Ashtekar, M.~Campiglia and A.~Henderson, Path
    Integrals and the WKB approximation in loop quantum cosmology,
    Phys. Rev. D\textbf{82} 124043 (2010);\\
    M.~Campiglia, A.~Henderson and W.~Nelson, Vertex expansion for the  
    Bianchi I model, Phys. Rev. D\textbf{82} 064036 (2010). 

\bibitem{ps} P.~Singh, Are loop quantum cosmologies never
    singular? Class. Quant. Grav. \textbf{26} 125005 (2009).

\bibitem{ambiguities} How probable is inflation? D.N. Page and S.W. Hawking, Nucl. Phys. B\textbf{298}, 789-809 (1988).

\bibitem{klm} Inflationary Theory and Alternative Cosmology, L.A. Kofman, A. Linde and V.F. Mukhanov, J. High Energy Phys. \textbf{10}, 057 (2002).

\bibitem{gt} G.W. Gibbons and N. Turok, The measure problem in cosmology, Phys. Rev. D\textbf{77}, 063516 (2008).

\bibitem{aads}A.~Ashtekar and D.~Sloan, Probability of inflation in
    loop quantum cosmology, Gen. Rel. Grav. 43, 3619-3656 (2011).

\bibitem{ck} A.~Corichi and A.~Karami, On the measure problem in slow roll inflation and loop quantum cosmology, Phys. Rev. D\textbf{83}, 104006 (2011). 

\bibitem{bl} L.~ Linsefors and A.~Barrau, Duration of inflation and conditions at the bounce as a prediction of effective isotropic loop quantum cosmology, Phys. Rev. D\textbf{87}, 123509 (2013).

\bibitem{akl} A.~Ashtekar, W.~Kaminski and J.~Lewandowski, Quantum
    field theory on a cosmological, quantum space-time, Physical
    Review D\textbf{79} 064030 (2009), arXiv:0901.0933.

\bibitem{rp1}R.~Penrose, \emph{Road to Reality}, sections 28.5 and 28.8 (Alfred A. Knopf, NY 2004).

\bibitem{aabg}A.~Ashtekar and B.~Gupt (in preparation)

\bibitem{ia} I.~Agullo Loop quantum cosmology, non-Gaussianity, and CMB power asymmetry, \texttt{ arXiv:1507.04703}.

\bibitem{isw1} L.~A. Kofman, A.~A.~Starobinskii, Effect of the cosmological constant on large-scale anisotropies in the microwave background, Sov. Astron. Lett. \textbf{11}, 271-274 (1985).

\bibitem{isw2} S.~Das and T.~Sourdeep, Suppressing CMB low multipoles with ISW effect, JCAP \textbf{1402}, 002 (2014). 

\bibitem{hkt} S.~A.~Hojman, K.~Kuchar and C.~Teitelboim,
    Geomtrodynamics regained, Ann. Phys. \textbf{96}, 88-135 (1976).

\bibitem{aan2} I. Agullo, A. Ashtekar and W. Nelson, An extension of
    the quantum theory of cosmological perturbations to the Planck
    era, \texttt{arXiv:1211.1354}, {\it Phys. Rev. D} {\bf 87}, 043507
    (2013).

\bibitem{bbcgk} A. Barrau, M. Bojowald, G. Calcagni, J. Grain, and M. Kagan, Anomaly-free cosmological perturbations
in effective canonical quantum gravity, JCAP \textbf{1505} (2015) 05, 051 

\bibitem{mcbg} J. Mielczarek, T. Cailleteau, A. Barrau, and J. Grain, Anomaly-free vector perturbations with holonomy corrections in loop quantum cosmology, Class.Quant.Grav. \textbf{29}, 085009 (2012).

\bibitem{cmbg} T. Cailleteau, J. Mielczarek, A. Barrau, and J. Grain, Anomaly-free scalar perturbations with holonomy corrections in loop quantum cosmology, Class.Quant.Grav. \textbf{29},  095010 (2012).

\bibitem{cbgv}  T. Cailleteau, A. Barrau, J. Grain, and F. Vidotto, Consistency of holonomy-corrected scalar, vector and tensor perturbations in Loop Quantum Cosmology, Phys. Rev. D \textbf{86}, 087301 (2012).

\bibitem{ed} E. Wilson-Ewing, Holonomy Corrections in the Effective Equations for Scalar Mode Perturbations in Loop Quantum Cosmology, Class.Quant.Grav. \textbf{29}, 085005 (2012).

\bibitem{bp} M. Bojowald and G.M. Paily, Deformed General Relativity and Effective Actions from Loop Quantum Gravity, Phys. Rev. D \textbf{86}, 104018 (2012).

\bibitem{mlb}  J. Mielczarek, L. Linsefors, and A. Barrau, Silent initial conditions for cosmological perturbations with a change of space-time signature, arXiv:1411.0272

\bibitem{bm} M. Bojowald and J. Mielczarek, Some implications of signature-change in cosmological models of loop quantum gravity, arXiv:1503.09154

\bibitem{lcbg} L. Linsefors, T. Cailleteau, A. Barrau, and J. Grain, Primordial tensor power spectrum in holonomy corrected Omega-LQC, Phys.Rev. D \textbf{87}, 107503 (2013).

\bibitem{bgslb} B.~Bolliet, J.~Grain, C.~Stahl, L.~Linsefors, A.~Barrau, Comparison of primordial tensor power spectra from the deformed algebra and dressed metric approaches in loop quantum cosmology, Phys. Rev. D\textbf{91} 084035 (2015). 

\bibitem{mb1} M.~Bojowald, {Absence of singularity in loop quantum
    cosmology}, Phys. Rev. Lett. \textbf{86} 5227-5230 (2001).

\bibitem{madrid} M.~Fernandez-Mendez, G.~A.~Mena
    Marugan, and J.~Olmedo,  Hybrid quantization of an inflationary
    universe, Phys.~Rev.~D~\textbf{86}, 024003 (2012).

\bibitem{bbbg}B.~Bonga and B.~Gupt (in preparation).

\bibitem{ana} I.~Agullo, W. Nelson and A.~Ashtekar, Preferred instantaneous vacuum for linear scalar fields in cosmological space-times, Phys. Rev. D\textbf{91}  064051 (2015).




\end{thebibliography}
\end{document}